# Polish and Silesian Non-Profit Organizations Liquidity Strategies[1]

**Grzegorz Michalski**[2] | *Wroclaw University of Economics, Wroclaw, Poland*
**Aleksander Mercik**[3] | *Wroclaw University of Economics, Wroclaw, Poland*

**Abstract**

The kind of realized mission inflows the sensitivity to risk. Among other factors, the risk results from decision about liquid assets investment level and liquid assets financing. The higher the risk exposure, the higher the level of liquid assets. If the specific risk exposure is smaller, the more aggressive could be the net liquid assets strategy. The organization choosing between various solutions in liquid assets needs to decide what level of risk is acceptable for her owners (or donors) and / or capital suppliers. The paper shows how, in authors opinion, decisions, about liquid assets management strategy inflow the risk of the organizations and its economical results during realization of main mission. Comparison of theoretical model with empirical data for over 450 Silesian nonprofit organization results suggests that nonprofit organization managing teams choose more risky aggressive liquid assets solutions than for-profit firms.

| **Keywords** | **JEL code** |
| --- | --- |
| Intrinsic liquidity value, nonprofit financial management, financial liquidity | G31, L31, M21 |

## INTRODUCTION

Organizations operate as taxed commercial businesses or non-taxed nonprofit organizations (Lane, 2001, pp. 1–17). As widely believed, the advantage of commercially driven businesses subsists in more effective management than in government controlled organizations (Nowicki, 2004, p. 29). In the paper we study the nonprofit organization liquid assets management. There is a group of organizations doing almost the same job as non-taxed government controlled organization, non-taxed nonprofit organization and taxed commercially managed business (Berger, 2008, pp. 46–47). That group of organizations face specific incumbent needs resulting in higher unemployment and other similar factors (Zietlow, 2010, pp. 238–248).

The main financial aim of the nonprofit organization (NPO) is not maximization of firm value but the best realization of the mission of that organization (Zietlow, 2007, pp. 6–7). But for assessment of financial decision of NPO analogous rules like for for-profit firms should be used (Brigham, 2006). One of that rules is the fact that higher risk is linked with higher cost of capital rate which should be used to evaluate the future results of decisions made by nonprofit organizations. This is also positively linked with the level of efficiency and effectiveness in realization of the NPO mission. Cost of financing net liquid assets depends on the risk included in the organization strategy of financing and / or investment in liquid assets.

---

[1] Acknowledgment: the research is financed from the Polish science budget resources in the years 2010–2012 as the research project NN113021139.
[2] E-mail: grzegorz.michalski@ue.wroc.pl.
[3] E-mail: mercik.aleksander@ue.wroc.pl.

Managing team in non-profit organizations have many good reasons for which their enterprises should possess some money resources reserves even if current interest rate is positive (Kim, 1998). The reasons may be classified into three main groups: the necessity of current expenses financing (transactional reason), fear of future cash flows uncertainty (precautional reason), future interest rate level uncertainty (speculative reason).

Liquid assets, especially cash, understood as money resources in organization safe are not a source of any or small interests. Maintaining liquidity reserve in the non-profit organization results from presumption that the value of lost income on account of interest will be recompensed by the benefits for incumbents of non-profit organization (Kim, 1998, Lee, 1990). The hypothetical benefits come from higher profitability that organization mission will be completed, thanks to adequate liquidity level. Then organizations maintaining such reserves assume that in equilibrium conditions, marginal liquidity value is equal to the interest rate of the Treasury Bonds investments (or interest rate being a cost of short-term credit we took out to obtain liquidity). Without doubt, the statement that investment in liquid assets does not bring any benefits and does not contribute to the realization of NPO mission may be rejected. From such a perspective, liquid assets would be treated as „necessary evil" linked only to the costs resulting from interests lost. Another incorrect conclusion would be an assumption that present net value always equals zero. It would be a result of the statement that due to the fact that marginal liquidity value is always equal to interests lost, cash reserves size has no significance at all (Henderson, 1989, p. 95, Kim, 1998, Lee, 1990, p. 540).

For organization being in possession of liquid reserves, the marginal utility of liquidity changes. Along with the growth in amount of cash possessed, the marginal cash value decreases. So it may be noticed that for the market Treasury Bond rate or short-term credit rate, it pays to keep some money reserve only to a certain level. There is a point corresponding with the optimal (critical) liquidity level, up to which the amount of liquid assets in the non-profit organization may be increased in a profit (Michalski, 2008b, Washam, 1989, p. 28, Henderson, 1989, Lee, 1990). The term: liquidity degree (or level) is connected with the known from economic literature conception of „liquidity container". The more liquid assets (which may be easily convertible into known amount of money resources and sensible only to a slight value change risk), the higher is enterprise liquidity level.

After exceeding this critical liquid assets level, the Treasury Bonds sale or taking out a short-term debt is unprofitable for the non-profit organization. The marginal benefit from higher cash reserve is lower than the cost of interests lost (Washam, 1989, Henderson, 1989).

In non-profit organization transactional and precautional liquid assets holdings on sufficient level allow for prompt fulfillment of internal (salary payments etc.) and external creditors (suppliers payment, etc.). The non-profit organization financial liquidity (operational and precautional) usually concerns operational activity and is not linked to investment activity. If it comes to enfeeblement or loss of operational and precautional liquidity in the non-profit organization, there is a menace (Scherr, 1989, Washam, 1989, Beck, 1993) of lowering decision making elasticity, deteriorating non-profit organization ability to set the organization mission, higher foreign capital raising cost, demobilization of donors, worsening non-profit organization position. In order to avoid such dangers, constant monitoring of non-profit organization financial liquidity is necessary as well as taking measures to guarantee its economic-financial equilibrium.

## 1 LIQUID ASSETS STRATEGIES AND COST OF FINANCING

Influence of liquid assets strategy on the rate of cost of capital financing non-profit organization and that influence on the economic results of NPO depend on relation between the kind of business risk taken by NPO, financial risk results from the financial leverage and individual risk characterizing the NPO. Capital providers take into consideration the nonprofit organization liquidity investment strategy while defining their claims as regards the rates of return. Restrictive strategy is perceived as more risky and therefore depending on investors risk aversion level, they tend to ascribe to the financed nonprofit organization applying restrictive strategy an additional expected risk premium. Ascribing the additional risk premium for applied liquidity investment strategy is reflected in the value of $β$ risk coefficient. For each strategy, the $β$ risk coefficient will be corrected by the corrective coefficient SZ corresponding to that specific strategy in relation to the current assets to cash revenues (CA / CR) situation.

Example: The risk free rate is 4 %, and rate of return on market portfolio is 18 %. If XYZ non-profit organization is a representative of W sector for which the non-leveraged risk coefficient $β_u = 0.77$. On the basis of Hamada relation (Hamada, 1972), we can estimate the equity cost rate that is financing the organization in case of each of the three strategies in the SZ1 variant.

$$\beta_l = \beta_u \times \left(1 + (1 - T) \times \frac{D}{E}\right) = 0.77 \times \left(1 + 0.81 \times \frac{0.4}{0.6}\right) = 1.19, \quad (1)$$

where: T — effective tax rate, here the assumption is taken that the NPO uses the tax-exempt debt and as a result there have about the same effective cost of debt as for profit-making organizations (Brigham, 2006, pp. 30–5, 7, 20),[4] D — organization financing capital coming from creditors (a sum of short term debt and long term debt $D = D_s + D_l$), E — organization financing capital coming from founders / owners of the organization, $β$ — risk coefficient, $β_u$ — risk coefficient for an assets of the non-profit organization that not use debt, $β_l$ — risk coefficient for an organization that applying the system of financing by creditors capital (here we have both asset and financial risk).

For restrictive strategy, where CA / CR is 0.3; the SZ risk premium is 0.2:

$$\beta_l^* = \beta_u \times \left(1 + (1 - T) \times \frac{D}{E}\right) \times (1 + SZ), \quad (2)$$

$$\beta_{lr}^* = 0.77 \times \left(1 + 0.81 \times \frac{0.4}{0.6}\right) \times 1.2 = 1.19 \times 1.2 = 1.43, \quad (3)$$

where: SZ — risk premium correction dependent on the liquidity investment strategy.

For moderate strategy, where CA / CR is 0.45 the SZ risk premium is 0.1:

$$\beta_{lm}^* = 0.77 \times \left(1 + 0.81 \times \frac{0.4}{0.6}\right) \times 1.1 = 1.19 \times 1.1 = 1.31. \quad (4)$$

For flexible strategy, where CA / CR is 0.6 the SZ risk premium is 0.01:

$$\beta_{lf}^* = 0.77 \times \left(1 + 0.81 \times \frac{0.4}{0.6}\right) \times 1.01 = 1.19 \times 1.01 = 1.2. \quad (5)$$

---

[4] According to Brigham (2006) even non-profit corporations that are exempt from taxation, and have the right to issue tax-exempt debt but individual contributions to these non-profit organizations can be deducted from taxable income by the donor, so: "non-profit businesses have access to tax-advantaged contributed capital".

Using that information we can calculate cost of equity rates for each liquidity investment strategy. For restrictive strategy:

$$k_{e_r} = \beta_l \times (k_m - k_{RF}) + k_{RF} = 1.43 \times 14\ \% + 4\ \% = 24\ \%. \tag{6}$$

For moderate strategy:

$$k_{e_m} = \beta_l \times (k_m - k_{RF}) + k_{RF} = 1.31 \times 14\ \% + 4\ \% = 22.3\ \%. \tag{7}$$

And for flexible strategy:

$$k_{e_f} = \beta_l \times (k_m - k_{RF}) + k_{RF} = 1.2 \times 14\ \% + 4\ \% = 20.8\ \%, \tag{8}$$

where: $k$ — rate of return expected by capital donors and at the same time (from nonprofit organization perspective) — cost of financing capital rate, $k_e$ — for cost rate of the equity, $k_{dl}$ — for long term debt rate, $k_{ds}$ — for short term debt rate, $k_m$ — for average rate of return on typical investment on the market, $k_{RF}$ — for risk free rate of return whose approximation is an average profitability of treasury bills in the country where the investment is made.

In similar way, we can calculate the risk premiums for XYZ alternative rates. We know that long term debt rates differ for $9\ \% \times (1 + SZ)$ in relation of equity to long term debt. From that we can get long term debt cost rates for each alternative strategy. For restrictive strategy:

$$k_{dl_r} = k_{e_r} - 9\ \% \times 1.2 = 24\ \% - 10.8\ \% = 13.2\ \%. \tag{9}$$

For moderate strategy:

$$k_{dl_m} = k_{e_m} - 9\ \% \times 1.1 = 22.3\ \% - 9.9\ \% = 12.4\ \%. \tag{10}$$

And for flexible strategy:

$$k_{dl_f} = k_{e_f} - 9\ \% \times 1.01 = 20.8\ \% - 9.1\ \% = 11.7\ \%. \tag{11}$$

Next we can calculate the risk premiums for XYZ alternative cost of short term rates. We know that short term debt rates differ for $12\ \% \times (1 + SZ)$ in relation of cost of equity rates to short term debt rates. From that we can get short term debt cost rates for each alternative strategy. For restrictive strategy:

$$k_{ds_r} = k_{e_r} - 12\ \% \times 1.2 = 24\ \% - 14.4\ \% = 9.6\ \%. \tag{12}$$

For moderate strategy:

$$k_{ds_m} = k_{e_m} - 12\ \% \times 1.1 = 22.3\ \% - 13.2\ \% = 9.1\ \%. \tag{13}$$

And for flexible strategy:

$$k_{ds_f} = k_{e_f} - 12\ \% \times 1.01 = 20.8\ \% - 12.1\ \% = 8.7\ \%. \tag{14}$$

As a result, cost of capital rate will amount to:

$$CC = \frac{E}{E+D_l+D_s} \times k_e + \frac{D_l}{E+D_l+D_s} \times k_{dl} \times (1-T) + \frac{D_s}{E+D_l+D_s} \times k_{ds} \times (1-T). \qquad (15)$$

However, for each strategy — this cost rate will be on another level (calculations in the Table 1).

**Table 1** Cost of capital and changes in economic results depending on the choice of liquidity investment strategy

| Liquidity investment strategy | Restrictive | Moderate | Flexible |
|---|---|---|---|
| Cash Revenues (CR) | 2 000 | 2 080 | 2 142.4 |
| Fixed assets (FA) | 1 400 | 1 445 | 1 480 |
| Current assets (CA) | 600 | 936 | 1 285 |
| Total assets (TA) = Total liabilities (TL) | 2 000 | 2 381 | 2 765 |
| Accounts payable (AP) | 300 | 468 | 643 |
| Capital invested ($E + D_l + D_s$) | 1 700 | 1 913 | 2 122 |
| Equity ($E$) | 680 | 765 | 849 |
| Long-term debt ($D_l$) | 340 | 383 | 424 |
| Short-term debt ($D_s$) | 680 | 765 | 849 |
| EBIT share in CR | 0.5 | 0.45 | 0.40 |
| Earnings before interests and taxes (EBIT)[5] | 1 000 | 936 | 857 |
| Free Cash Flows in 1 to n periods ($FCF_{1..n}$) | 1 000 | 936 | 857 |
| Initial Free Cash Flows in year 0 ($FCF_o$) | −1 700 | −1 913 | −2 122 |
| SZ risk premium correction | 0.2 | 0.1 | 0.01 |
| Leveraged and corrected risk coefficient $\beta_l$ | 1.428 | 1.309 | 1.2019 |
| Cost of equity rate ($k_e$) | 23.99 % | 22.33 % | 20.83 % |
| Long-term debt rate ($k_{dl}$) | 13.19 % | 12.43 % | 11.74 % |
| Short-term debt rate ($k_{ds}$) | 9.59 % | 9.13 % | 8.71 % |
| Cost of capital (CC) | 14.84 % | 13.90 % | 13.05 % |
| **Economic result of liquidity strategy** | 5 037.77 | 4 821.18 | 4 443.17 |

Source: Hypothetical data

As shown in the Table 1, rates of the cost of capital financing of non-profit organization are different due to different approaches to liquidity investment. The lowest rate: $CC = 13.1\%$; is observed in flexible strategy because that strategy is linked with the smallest level of risk but the highest economic effect is bound to restrictive strategy of investment in liquidity.

Cost of capital for restrictive strategy of investment in liquidity:

$$CC_r = \frac{680}{1\,700} \times 24\% + \frac{340}{1\,700} \times 13.2\% \times (1 - 0.19) + \frac{680}{1\,700} \times 9.6\% \times (1 - 0.19) = 14.8\%. \qquad (16)$$

Expected growth of economic result of liquidity strategy:

$$\Delta ER_r = FCF_o + \frac{FCF_{1...n}}{CC} = -1\,700 + \frac{1\,000}{0.148} = 5\,057. \qquad (17)$$

---

[5] Because of exempt of taxation, EBIT is equal to net operating profit after taxes (NOPAT).

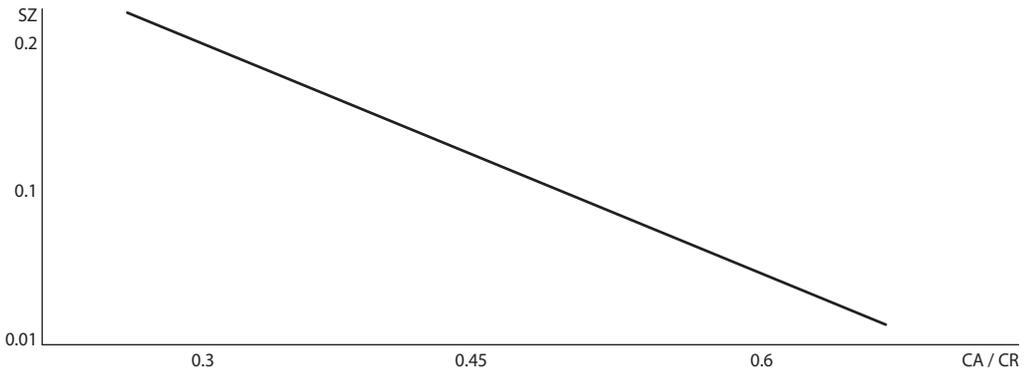

**Figure 1** The shape of line of correction SZ as a function of CA / CR in the SZ1 variant

**Source:** Hypothetical data

Cost of capital for moderate strategy of investment in liquidity:

$$CC_m = \frac{765}{1\,913} \times 22.3\% + \frac{383}{1\,913} \times 12.4\% \times (1-0.19) + \frac{765}{1\,913} \times 9.1\% \times (1-0.19) = 13.9\%. \qquad (18)$$

Expected growth of economic result for that strategy:

$$\Delta ER_m = -1\,913 + \frac{936}{0.139} = 4\,821. \qquad (19)$$

Cost of capital for flexible strategy of investment in liquidity:

$$CC_f = \frac{849}{2\,122} \times 20.8\% + \frac{424}{2\,122} \times 11.7\% \times (1-0.19) + \frac{849}{2\,122} \times 8.7\% \times (1-0.19) = 13.1\%. \qquad (20)$$

Expected growth of economic result for flexible strategy:

$$\Delta ER_f = -2\,122 + \frac{857}{0.131} = 4\,420. \qquad (21)$$

The expected after crisis changes will correct both the market liquidity value and the cost of capital rate. Both factors influence the target (and optimal) liquidity level for nonprofit organization. That will result with more restrictive liquidity levels because of change in equilibrium point for intrinsic and market liquidity values (Michalski, 2010, Golawska-Witkowska, 2006, p. 144, Jaworski, 2010, pp. 366–368). The cost of capital will be higher after crisis than before (Fernandez, 2011, pp. 4–7, Fernandez, 2010, pp. 4–7, Fernandez, 2008, pp. 5–8). That will result in changes in efficiency of liquidity policy for nonprofit organizations (as shown in the Table 2).

**Table 2** Cost of capital and changes in economic results depending on the choice of liquidity investment strategy

| Liquidity investment strategy | Restrictive | Moderate | Flexible |
|---|---|---|---|
| Cash Revenues (CR) | 2 000 | 2 080 | 2 142.4 |
| Fixed assets (FA) | 1 400 | 1 445 | 1 480 |
| Current assets (CA) | 600 | 936 | 1 285 |
| Total assets (TA) = Total liabilities (TL) | 2 000 | 2 381 | 2 765 |
| Accounts payable (AP) | 300 | 468 | 643 |
| Capital invested ($E + D_l + D_s$) | 1 700 | 1 913 | 2 122 |
| Equity ($E$) | 680 | 765 | 849 |
| Long-term debt ($D_l$) | 340 | 383 | 424 |
| Short-term debt ($D_s$) | 680 | 765 | 849 |
| EBIT share in CR | 0.5 | 0.45 | 0.40 |
| Earnings before interests and taxes (EBIT) | 1 000 | 936 | 857 |
| Free Cash Flows in 1 to n periods ($FCF_{1..n}$) | 1 000 | 936 | 857 |
| Initial Free Cash Flows in year 0 ($FCF_0$) | −1 700 | −1 913 | −2 122 |
| SZ risk premium correction | 0.2 | 0.1 | 0.01 |
| Leveraged and corrected risk coefficient $β_l$ | 1.428 | 1.309 | 1.2019 |
| Cost of equity rate ($k_e$) | 27.85 % | 25.94 % | 24.23 % |
| Long-term debt rate ($k_{dl}$) | 17.05 % | 16.04 % | 15.14 % |
| Short-term debt rate ($k_{ds}$) | 13.45 % | 12.74 % | 12.11 % |
| Cost of capital (CC) | 18.26 % | 17.10 % | 16.07 % |
| **Economic result of liquidity strategy** | 3 777 | 3 559.18 | 3 211.06 |

**Source:** Hypothetical data

As shown in the Table 2, the after crisis changes influence the efficiency of the liquidity investment of nonprofit organization. It is natural that changes depend on NPO risk sensitivity. Depending on their risk sensitivity, an additional risk premium for an NPO that implemented this type of strategy should be used. As presented in the Figure 2, we have stronger risk sensitivity than in previous situation.

**Figure 2** The shape of line of correction SZ as a function of CA / CR in the SZ2 variant

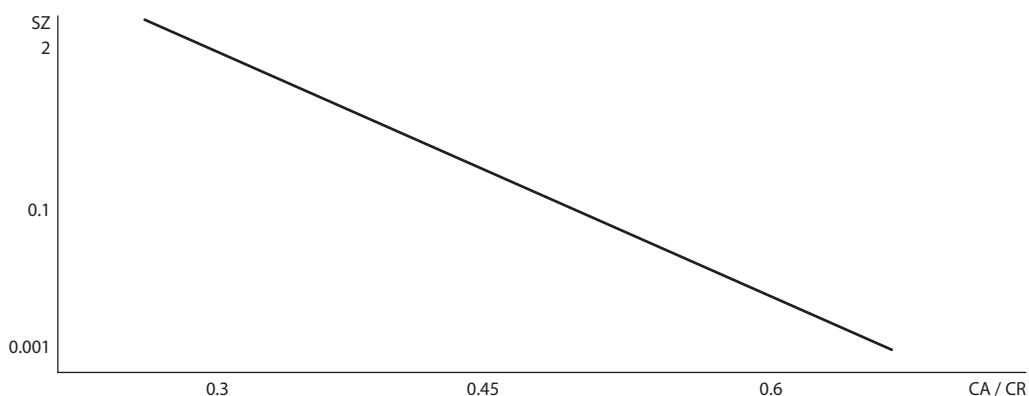

**Source:** Hypothetical data

In the Table 3 there are calculations for that variant. For each strategy the cost of capital rate *CC* will be on another level.

Table 3 Cost of capital and changes in economic results depending on the choice of liquidity investment strategy (before the crisis influence)

| Liquidity investment strategy | Restrictive | Moderate | Flexible |
|---|---|---|---|
| Cash Revenues (CR) | 2 000 | 2 080 | 2 142.4 |
| Fixed assets (FA) | 1 400 | 1 445 | 1 480 |
| Current assets (CA) | 600 | 936 | 1 285 |
| Total assets (TA) = Total liabilities (TL) | 2 000 | 2 381 | 2 765 |
| Accounts payable (AP) | 300 | 468 | 643 |
| Capital invested $(E + D_l + D_s)$ | 1 700 | 1 913 | 2 122 |
| Equity (E) | 680 | 765 | 849 |
| Long-term debt $(D_l)$ | 340 | 383 | 424 |
| Short-term debt $(D_s)$ | 680 | 765 | 849 |
| EBIT share in CR | 0.5 | 0.45 | 0.40 |
| Earnings before interests and taxes (EBIT) | 1 000 | 936 | 857 |
| Free Cash Flows in 1 to n periods $(FCF_{1..n})$ | 1 000 | 936 | 857 |
| Initial Free Cash Flows in year 0 $(FCF_o)$ | –1 700 | –1 913 | –2 122 |
| SZ risk premium correction | 2 | 0.1 | 0.001 |
| Leveraged and corrected risk coefficient $\beta_l$ | 3.5574 | 1.30438 | 1.186986 |
| Cost of equity rate $(k_e)$ | 53.80 % | 22.26 % | 20.62 % |
| Long-term debt rate $(k_{dl})$ | 26.80 % | 12.36 % | 11.61 % |
| Short-term debt rate $(k_{ds})$ | 17.80 % | 9.06 % | 8.61 % |
| Cost of capital (CC) | 31.63 % | 13.84 % | 12.92 % |
| **Economic result of liquidity strategy** | 1 461 | 4 849 | 4 513 |

Source: Hypothetical data

In similar way we can calculate for situation with higher after crisis cost of capital rates levels. The result is presented in the Table 4.

Figure 3 The shape of line of correction SZ as a function of CA / CR in the SZ3 variant

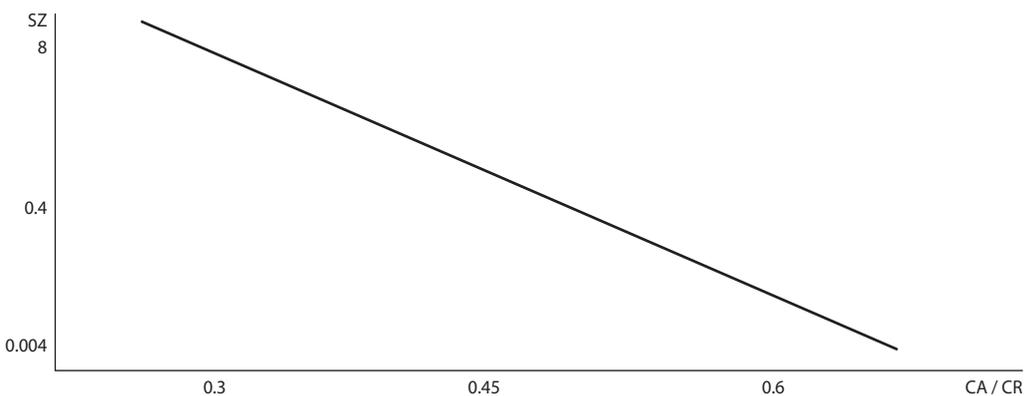

Source: Hypothetical data

Table 4 Cost of capital and changes in economic results depending on the choice of liquidity investment strategy

| Liquidity investment strategy | Restrictive | Moderate | Flexible |
|---|---|---|---|
| Cash Revenues (CR) | 2 000 | 2 080 | 2 142.4 |
| Fixed assets (FA) | 1 400 | 1 445 | 1 480 |
| Current assets (CA) | 600 | 936 | 1 285 |
| Total assets (TA) = Total liabilities (TL) | 2 000 | 2 381 | 2 765 |
| Accounts payable (AP) | 300 | 468 | 643 |
| Capital invested ($E + D_l + D_s$) | 1 700 | 1 913 | 2 122 |
| Equity (E) | 680 | 765 | 849 |
| Long-term debt ($D_l$) | 340 | 383 | 424 |
| Short-term debt ($D_s$) | 680 | 765 | 849 |
| EBIT share in CR | 0.5 | 0.45 | 0.40 |
| Earnings before interests and taxes (EBIT) | 1 000 | 936 | 857 |
| Free Cash Flows in 1 to n periods ($FCF_{1..n}$) | 1 000 | 936 | 857 |
| Initial Free Cash Flows in year 0 ($FCF_o$) | –1 700 | –1 913 | –2 122 |
| SZ risk premium correction | 2 | 0.1 | 0.001 |
| Leveraged and corrected risk coefficient $\beta_l$ | 3.5574 | 1.30438 | 1.186986 |
| Cost of equity rate ($k_e$) | 61.92 % | 25.87 % | 23.99 % |
| Long-term debt rate ($k_{dl}$) | 34.92 % | 15.97 % | 14.98 % |
| Short-term debt rate ($k_{ds}$) | 25.92 % | 12.67 % | 11.98 % |
| Cost of capital (CC) | 38.82 % | 17.04 % | 15.91 % |
| **Economic result of liquidity strategy** | 877 | 3 580 | 3 266 |

**Source:** Hypothetical data

## 2 EMPIRICAL DATA FOR POLAND

Data collected about Polish NPO show their liquidity strategies for 2009 and 2010. If we compare it with the results of profit-oriented Polish organizations we can conclude that the average length of operating cycle and net operating cycle (cash cycle) is shorter than for average for profit organizations. Observation of NPO data can inform us about interesting customs of NPO managing teams. Generally, basing on the data collected from Opolskie area in Poland, for 2009 and 2010 years, we can see that average operating cycle for such group of organizations vary differ, in 2009 was short (about 5.89 days for 2009 data, with standard deviation = SD = 22.69 days) and in 2010 was shorter (about 3.59 days for 2010 data, with SD = 9.35 days).

Table 5 Operating cycle indicators for Opolskie nonprofit organizations in 2009 and 2010

|  | Operating cycle | Cash cycle | ROA | ROE |
|---|---|---|---|---|
| M 2009 | 5.89 | –1.47 | –169.96 % | 7.15 % |
| SD 2009 | 22.69 | 33.55 | 1 272.09 % | 533.11 % |
| M 2010 | 3.59 | –7.1 | 2.21 % | 1 258.21 % |
| SD 2010 | 9.35 | 50.34 | 120.35 % | 11 463.45 % |

**Note:** SD = standard deviation, M = arithmetic mean.
**Source:** Own calculation for 80 selected nonprofits in OPOLSKIE (Bopp, 2011)

Selected data shows that there is no hard link between operating cycle and ROA and ROE results. Operating cycle policy must be first of all a slave of the best realization of the mission nonprofit organization. The economic results are important, but the second or even third in the queue of the aims.

Table 6 Liquid assets indicators for Polish nonprofit organizations in 2009 and 2010

| 2009 | CR | Assets | CA | Current Ratio | Quick Ratio | Cash Ratio | INV |
|---|---|---|---|---|---|---|---|
| Number of observations | 2 283 | 2 292 | 2 294 | 1 473 | 1 471 | 1 467 | 2 291 |
| Mean | 483 699 | 834 187 | 201 034 | 1 092 | 526 | 474 | 6 284 |
| SD | 1 636 492 | 13 073 895 | 1 315 942 | 23 069 | 5 201 | 4 998 | 46 105 |
| Median | 76 979 | 24 732 | 19 062 | 5.6 | 5.42 | 4.54 | – |
| Winsorized mean | 693 825 | 352 948 | 172 751 | 63 | 62 | 56.3 | – |
| Truncated (trimmed) mean | 141 493 | 58 492 | 34 793 | 12 | 12 | 10.21 | – |

| 2010 | AR | Cash equivalents | E | Dl | Ds | ROA | ROE |
|---|---|---|---|---|---|---|---|
| Number of observations | 2 290 | 2 292 | 2 294 | 2 293 | 2 293 | 2 266 | 2 247 |
| Mean | 32 043 | 172 066 | 688 121 | 11 026 | 47 152 | –0.57 | –0.04 |
| SD | 605 949 | 1 291 873 | 12 967 335 | 112 797 | 312 128 | 23 | 23 |
| Median | – | 13 902 | 17 037 | – | 607 | 0 | 0.30 |
| Winsorized mean | 11 318 | 116 842 | 207 907 | – | 35 605 | 1 | 1 |
| Truncated (trimmed) mean | 2 282 | 25 330 | 37 026 | – | 6 822 | 0 | 0.31 |

**Note:** SD = standard deviation, M = arithmetic mean, AR = accounts receivable, E = fund capital, $D_l$ = long-term debt, $D_s$ = short-term debt, INV = inventories.
**Source:** Own calculation for 1000+ selected nonprofits in Poland (Bopp, 2011)

According to data received from 1000+ Polish NPOs, the average NPO investment in liquid assets is more aggressive than in profit-oriented organizations. Average Polish NPO accounts receivable period for 2009–2010 data is about 23 days (5.8 days using winsorized mean and 5.8 days using truncated mean). Average Polish for profit accounts receivable period for 2009–2010 data is about 46 days (Dudycz, 2011). Average Polish NPO inventory period for 2009–2010 data is about 4.7 days. Average Polish for profit inventory period for 2009–2010 data is about 39 days.

The above observation suggests that in case of Polish NPO we have figure 6 situation. Is it small risk exposition or rather smaller aversion of managing teams? Unfortunately rather the second.

## 3 SILESIAN EMPIRICAL DATA

Distribution analysis commands to understand the financial management process in Silesian NPO. Probability distribution function and statistical dispersion of financial data could provide valuable information about current financial conditions in not-for-profit businesses.

An important aspect is the shape of a distribution, showing the frequency of values from different ranges of the variable. The analysis of all financial ratios produced some interesting results. Skewness (a measure of the asymmetry of the probability distribution) is clearly different from 0, which means that distribution is asymmetrical. Boxplots (the Figures 4 and 5) testify that all analyzed data are not normally distributed. Especially a boxplot is a convenient way of graphically depicting groups of numerical data through their summaries: minimum, lower quartile (Q1), median, upper quartile (Q3), maximum. The location of the box within the whiskers provides an insight into the asymmetry of the sample's distribution. The samples are extremely positively skewed. A thinner box relative to the whiskers indicates a thinner peak.

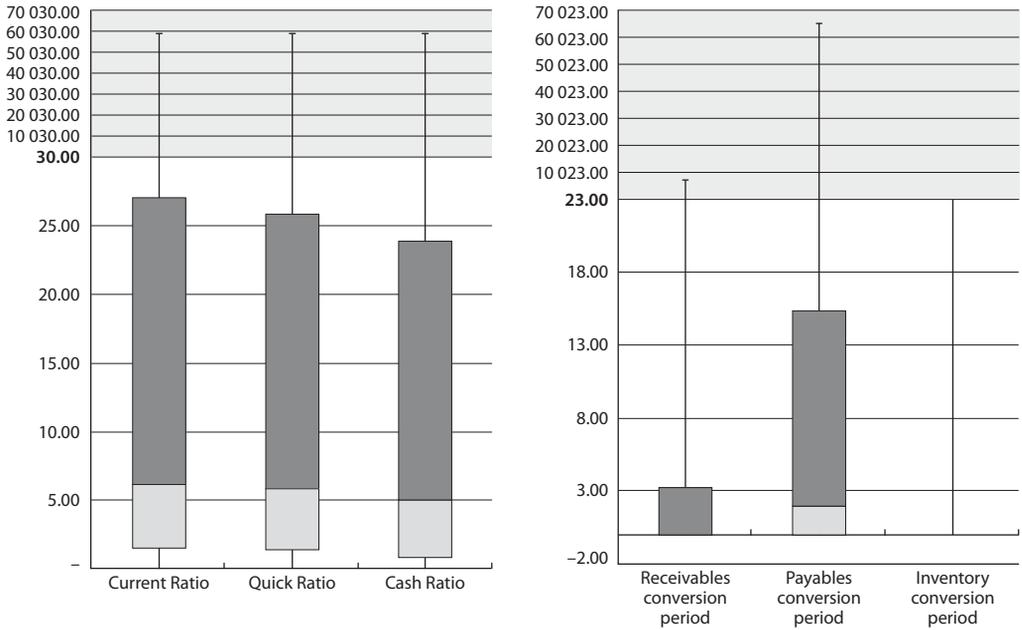

**Figure 4** Boxplots — liquid ratios nad conversion periods (2009)

**Source:** Own calculation for over 450 selected nonprofits in Silesia (Bopp, 2011)

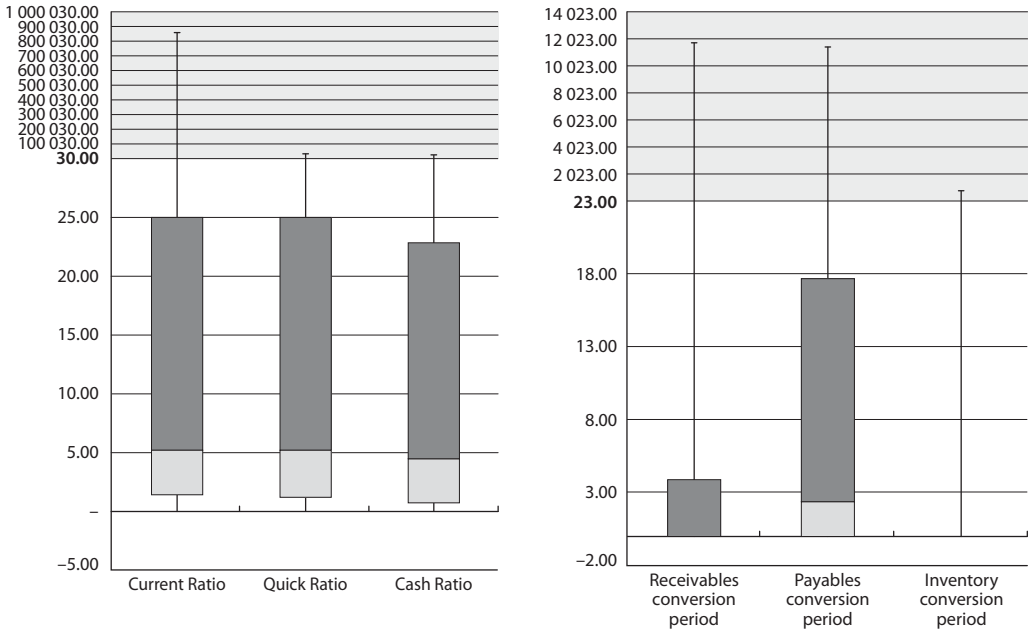

**Figure 5** Boxplots — liquid ratios and conversion periods (2010)

**Source:** Own calculation for over 450 selected nonprofits in Silesia (Bopp, 2011)

Table 7 Liquid assets indicators for Silesian nonprofit organizations in 2009 and 2010

| 2009 | Receivables conversion period | Payables conversion period | Inventory conversion period | Current Ratio | Quick Ratio | Cash Ratio |
|---|---|---|---|---|---|---|
| Size of population | 707 | 709 | 708 | 449 | 448 | 448 |
| Average | 124.95 | 1 171.96 | 7.64 | 370.14 | 369.40 | 296.87 |
| Standard deviation | 2 755.26 | 24 858.80 | 82.81 | 3 405.82 | 3 408.52 | 3 220.69 |
| Median | 0.00 | 2.00 | 0.00 | 6.26 | 5.99 | 4.98 |
| Truncated mean | 2.58 | 8.86 | – | 12.41 | 12.06 | 10.95 |
| Winsorized mean | 8.56 | 31.86 | 0.00 | 71.08 | 71.14 | 62.60 |
| Skewness | 26.29 | 25.49 | 17.37 | 13.81 | 13.80 | 15.65 |
| Maximum | 73 000.00 | 651 462.75 | 1 724.32 | 58 415.22 | 58 415.22 | 58 396.28 |
| Minimum | 0.00 | 0.00 | 0.00 | 0.00 | 0.00 | 0.00 |

| 2010 | Receivables conversion period | Payables conversion period | Inventory conversion period | Current Ratio | Quick Ratio | Cash Ratio |
|---|---|---|---|---|---|---|
| Size of population | 711 | 712 | 711 | 459 | 458 | 457 |
| Average | 58.47 | 89.53 | 6.02 | 1 980.06 | 167.24 | 156.68 |
| Standard deviation | 604.86 | 692.38 | 46.55 | 40 314.73 | 1 627.38 | 1 575.16 |
| Median | 0.00 | 2.33 | 0.00 | 5.24 | 5.19 | 4.48 |
| Truncated mean | 2.99 | 9.92 | – | 11.65 | 11.45 | 9.87 |
| Winsorized mean | 11.31 | 39.66 | 0.00 | 62.60 | 62.77 | 55.70 |
| Skewness | 14.72 | 12.39 | 11.05 | 21.42 | 18.88 | 18.96 |
| Maximum | 11 643.87 | 11 360.33 | 769.10 | 863 747.33 | 33 331.90 | 32 281.67 |
| Minimum | 0.00 | 0.00 | 0.00 | 0.00 | 0.00 | –0.07 |

Source: Own calculation for over 450 selected nonprofits in Silesia (Bopp, 2011).

The right side tail of the probability density function is much longer than the left side. The mean (and standard deviation) can be heavily influenced by extreme values in the tails of a variable. In this case a truncated mean and a Winsorized mean are more useful estimators. Comparing to the mean, they are less sensitive to outliers than the mean (Heilpern, 1999) but it still gives a reasonable estimate of central tendency. Truncated mean rejects some parts of the data from the top or from the bottom end, (typically an equal amount at each end) and then calculate the arithmetic mean of the remaining data (Rothenberg, 1966). On the other hand, a Winsorized mean involves the calculation of the mean after replacing given parts of a probability distribution or sample at the high and low end with the most extreme remaining values (Wilcox, 2003).

Data presented in the Table 8 show the values of liquid assets indicators for specific mission. Most of the analyzed NPOs can qualify for several "sectors". For this reason data from the same organizations are included in several tables.

**Table 8** Liquid assets indicators for Silesian nonprofit organizations (social assistance, including assistance to families and individuals in difficult situations, and equalization of opportunities for those families and individuals) in 2009 and 2010

| 2009–2010 | Receivables conversion period | Payables conversion period | Inventory conversion period | Current Ratio | Quick Ratio | Cash Ratio |
|---|---|---|---|---|---|---|
| Size of population | 610 | 610 | 610 | 409 | 409 | 409 |
| Average | 11.26 | 1 303.49 | 11.24 | 2 433.75 | 399.51 | 328.69 |
| Standard deviation | 79.23 | 26 759.15 | 98.08 | 42 810.49 | 3 557.42 | 3 348.06 |
| Median | – | 2.22 | – | 6.74 | 6.41 | 5.42 |
| Truncated mean | 2.14 | 7.76 | – | 14.46 | 14.13 | 12.96 |
| Winsorized mean | 7.36 | 30.43 | 0.00 | 82.80 | 82.56 | 81.41 |
| Skewness | 17.52 | 23.75 | 13.16 | 20.06 | 13.27 | 15.06 |
| Maximum | 1 718 | 651 463 | 1 724 | 863 747 | 58 415 | 58 396 |
| Minimum | 0 | 0 | 0 | 0 | 0 | 0 |

**Source:** Own calculation for over 450 selected nonprofits in Silesia (Bopp, 2011)

**Table 9** Liquid assets indicators for Silesian nonprofit organizations (activities for the integration and professional and social reintegration of those at risk of social exclusion) in 2009 and 2010

| 2009–2010 | Receivables conversion period | Payables conversion period | Inventory conversion period | Current Ratio | Quick Ratio | Cash Ratio |
|---|---|---|---|---|---|---|
| Size of population | 463 | 463 | 463 | 303 | 303 | 303 |
| Average | 12.72 | 1 442.49 | 12.67 | 3 219.17 | 472.51 | 387.33 |
| Standard deviation | 87.24 | 30 274.87 | 109.70 | 49 731.35 | 4 080.96 | 3 856.45 |
| Median | – | 2.34 | – | 10.09 | 8.77 | 7.45 |
| Truncated mean | 1.78 | 7.84 | – | 15.44 | 15.03 | 13.47 |
| Winsorized mean | 6.46 | 26.88 | – | 80.66 | 80.37 | 69.75 |
| Skewness | 16.72 | 21.52 | 12.21 | 17.27 | 11.80 | 13.26 |
| Maximum | 1 718 | 651 463 | 1 724 | 863 747 | 58 415 | 58 396 |
| Minimum | 0 | 0 | 0 | 0 | 0 | 0 |

**Source:** Own calculation for over 450 selected nonprofits in Silesia (Bopp, 2011)

**Table 10** Liquid assets indicators for Silesian nonprofit organizations (charitable activities) in 2009 and 2010

| 2009–2010 | Receivables conversion period | Payables conversion period | Inventory conversion period | Current Ratio | Quick Ratio | Cash Ratio |
|---|---|---|---|---|---|---|
| Size of population | 264 | 264 | 264 | 156 | 156 | 156 |
| Average | 288.54 | 34.59 | 8.21 | 134.80 | 134.38 | 114.64 |
| Standard deviation | 4 492.34 | 165.71 | 40.94 | 887.26 | 887.32 | 758.27 |
| Median | – | 1.48 | – | 6.45 | 5.72 | 4.58 |
| Truncated mean | 3.50 | 8.06 | – | 10.60 | 10.00 | 8.65 |
| Winsorized mean | 10.28 | 25.67 | – | 47.13 | 46.15 | 44.97 |
| Skewness | 16.25 | 9.39 | 6.24 | 10.62 | 10.62 | 10.36 |
| Maximum | 73 000 | 2 030 | 365 | 10 463 | 10 463 | 8 815 |
| Minimum | 0 | 0 | 0 | 0 | 0 | 0 |

**Source:** Own calculation for over 450 selected nonprofits in Silesia (Bopp, 2011)

Table 11 Liquid assets indicators for Silesian nonprofit organizations (activities for national and ethnic minorities and regional language) in 2009 and 2010

| 2009–2010 | Receivables conversion period | Payables conversion period | Inventory conversion period | Current Ratio | Quick Ratio | Cash Ratio |
|---|---|---|---|---|---|---|
| Size of population | 548 | 548 | 548 | 346 | 346 | 346 |
| Average | 166.89 | 1 520.99 | 10.98 | 2 851.70 | 446.75 | 375.91 |
| Standard deviation | 3 138.39 | 28 274.52 | 102.15 | 46 541.48 | 3 845.26 | 3 633.77 |
| Median | – | 1.52 | – | 7.39 | 7.32 | 6.12 |
| Truncated mean | 2.22 | 7.63 | – | 16.89 | 16.78 | 15.29 |
| Winsorized mean | 7.48 | 28.76 | – | 96.43 | 95.35 | 92.48 |
| Skewness | 22.96 | 22.40 | 12.92 | 18.46 | 12.37 | 13.89 |
| Maximum | 73 000 | 651 463 | 1 724 | 863 747 | 58 415 | 58 396 |
| Minimum | 0 | 0 | 0 | 0 | 0 | 0 |

**Source:** Own calculation for over 450 selected nonprofits in Silesia (Bopp, 2011)

Table 12 Liquid assets indicators for Silesian nonprofit organizations (health protection and promotion) in 2009 and 2010

| 2009–2010 | Receivables conversion period | Payables conversion period | Inventory conversion period | Current Ratio | Quick Ratio | Cash Ratio |
|---|---|---|---|---|---|---|
| Size of population | 530 | 530 | 530 | 358 | 358 | 358 |
| Average | 147.58 | 1 495.75 | 12.10 | 2 873.91 | 549.22 | 464.82 |
| Standard deviation | 3 170.71 | 28 706.30 | 101.46 | 45 784.82 | 4 185.91 | 3 978.99 |
| Median | – | 2.73 | – | 7.42 | 6.71 | 5.79 |
| Truncated mean | 2.61 | 8.96 | – | 14.47 | 14.18 | 12.67 |
| Winsorized mean | 10.24 | 32.31 | – | 85.80 | 85.60 | 77.94 |
| Skewness | 23.02 | 22.13 | 13.00 | 18.73 | 10.58 | 11.74 |
| Maximum | 73 000 | 651 463 | 1 724 | 863 747 | 58 415 | 58 396 |
| Minimum | 0 | 0 | 0 | 0.01 | 0.0014 | 0 |

**Source:** Own calculation for over 450 selected nonprofits in Silesia (Bopp, 2011)

Table 13 Liquid assets indicators for Silesian nonprofit organizations (support economic development activities, including the development of entrepreneurship) in 2009 and 2010

| 2009–2010 | Receivables conversion period | Payables conversion period | Inventory conversion period | Current Ratio | Quick Ratio | Cash Ratio |
|---|---|---|---|---|---|---|
| Size of population | 684 | 684 | 684 | 424 | 424 | 424 |
| Average | 138.93 | 1 245.81 | 6.69 | 2 290.32 | 329.67 | 314.45 |
| Standard deviation | 2 815.96 | 25 311.65 | 57.82 | 42 036.50 | 3 331.97 | 3 295.33 |
| Median | – | 1.90 | – | 6.30 | 5.96 | 5.24 |
| Truncated mean | 2.67 | 8.91 | – | 12.45 | 12.18 | 10.80 |
| Winsorized mean | 9.63 | 32.52 | – | 60.87 | 60.68 | 55.26 |
| Skewness | 25.48 | 25.02 | 15.57 | 20.45 | 14.99 | 15.25 |
| Maximum | 73 000 | 651 463 | 1 222 | 863 747 | 58 415 | 58 396 |
| Minimum | 0 | 0 | 0 | 0 | 0 | 0 |

**Source:** Own calculation for over 450 selected nonprofits in Silesia (Bopp, 2011)

**Table 14** Liquid assets indicators for Silesian nonprofit organizations (science, education, higher education) in 2009 and 2010

| 2009–2010 | Receivables conversion period | Payables conversion period | Inventory conversion period | Current Ratio | Quick Ratio | Cash Ratio |
|---|---|---|---|---|---|---|
| Size of population | 560 | 560 | 560 | 347 | 347 | 347 |
| Average | 205.84 | 308.22 | 7.73 | 2 573.31 | 177.24 | 162.98 |
| Standard deviation | 3 159.28 | 4 808.22 | 67.65 | 46 367.95 | 1 886.03 | 1 806.79 |
| Median | – | 2.11 | – | 3.52 | 3.44 | 2.60 |
| Truncated mean | 3.05 | 11.07 | – | 7.37 | 7.13 | 6.03 |
| Winsorized mean | 10.11 | 39.89 | – | 40.95 | 40.78 | 34.02 |
| Skewness | 22.09 | 22.73 | 13.84 | 18.62 | 16.28 | 16.67 |
| Maximum | 73 000 | 112 290 | 1 222 | 863 747 | 33 332 | 32 282 |
| Minimum | 0 | 0 | 0 | 0 | 0 | 0 |

**Source:** Own calculation for over 450 selected nonprofits in Silesia (Bopp, 2011)

**Table 15** Liquid assets indicators for Silesian nonprofit organizations (rescue and civil protection) in 2009 and 2010

| 2009–2010 | Receivables conversion period | Payables conversion period | Inventory conversion period | Current Ratio | Quick Ratio | Cash Ratio |
|---|---|---|---|---|---|---|
| Size of population | 309 | 309 | 309 | 221 | 221 | 221 |
| Average | 7.56 | 2 151.28 | 6.60 | 4 369.62 | 610.76 | 501.47 |
| Standard deviation | 27.76 | 37 058.53 | 74.63 | 58 225.75 | 4 786.83 | 4 512.45 |
| Median | – | 3.52 | – | 5.73 | 5.59 | 5.43 |
| Truncated mean | 2.20 | 8.10 | – | 13.38 | 13.25 | 12.32 |
| Winsorized mean | 7.74 | 29.21 | – | 74.79 | 74.70 | 73.98 |
| Skewness | 6.63 | 17.58 | 14.88 | 14.75 | 9.99 | 11.32 |
| Maximum | 286 | 651 463 | 1 222 | 863 747 | 58 415 | 58 396 |
| Minimum | 0 | 0 | 0 | 0 | 0 | 0 |

**Source:** Own calculation for over 450 selected nonprofits in Silesia (Bopp, 2011)

**Table 16** Liquid assets indicators for Silesian nonprofit organizations (other sectors) in 2009 and 2010

| 2009–2010 | Receivables conversion period | Payables conversion period | Inventory conversion period | Current Ratio | Quick Ratio | Cash Ratio |
|---|---|---|---|---|---|---|
| Size of population | 813 | 813 | 813 | 507 | 507 | 507 |
| Average | 121.65 | 1 051.93 | 5.86 | 1 967.92 | 327.80 | 270.45 |
| Standard deviation | 2 584.79 | 23 218.35 | 51.84 | 38 453.58 | 3 196.00 | 3 010.05 |
| Median | – | 1.66 | – | 6.37 | 6.20 | 5.18 |
| Truncated mean | 2.38 | 8.10 | – | 12.73 | 12.32 | 10.86 |
| Winsorized mean | 9.71 | 30.37 | – | 64.02 | 63.53 | 56.50 |
| Skewness | 27.72 | 27.28 | 17.82 | 22.34 | 14.81 | 16.75 |
| Maximum | 73 000 | 651 463 | 1 222 | 863 747 | 58 415 | 58 396 |
| Minimum | 0 | 0 | 0 | 0 | 0 | 0 |

**Source:** Own calculation for over 450 selected nonprofits in Silesia (Bopp, 2011)

**CONCLUSION**

As shown in our findings, depending on kind of realized mission and sensitivity to risk NPOs should chose liquid assets investment level and corresponding (based on liquid assets) financing. The kind of organization influences the best strategy choice. If a risk exposure is greater, the higher level of inventories, accounts receivable and operating cash should occur. (Michalski, 2008a). If the risk exposure is smaller, the more aggressive will be the net liquid assets strategy and smaller level of inventories. The organization choosing between various solutions in liquid assets needs to decide what level of risk is acceptable for its owners and capital suppliers. That choice results in financing consequences, especially in cost level. It is a basis for considerations about relations between risk and expected benefits from the liquid assets decision and its impact on financing costs for both nonprofit or profit-oriented organizations. Decisions on liquid assets management strategy and choice between kind of taxed or non-taxed form inflow the risk of the organizations and its economical results during realization of main mission. Comparing theoretical model with empirical data for over 450 Silesian nonprofit organization result, suggests that nonprofit organization managing teams choose more risky aggressive liquid assets solutions than for-profit firms. That observation suggests that here, in Silesian NPO case we have the Figure 1 situation with smallest risk exposure solution in managing team mind. But in fact probably there is not a smaller risk exposition but rather smaller aversion of managing teams of Silesian NPOs.